\documentclass[a4paper,aps,preprintnumbers,showpacs,amsmath,amssymb,prl]{revtex4}
\usepackage{mathrsfs}
\usepackage{amsmath}
\usepackage[dvips]{graphicx}
\usepackage{epsfig}
\usepackage[dvips]{graphics,graphicx}
\begin{document}

\title{Two-photon nonlinear spectroscopy of periodically trapped ultracold atoms in a cavity}

\author{Tarun Kumar$^{1}$, Aranya B.\ Bhattacherjee$^{2,3}$ and ManMohan$^{1}$}

\address{$^{1}$Department of Physics and Astrophysics, University of Delhi, Delhi-110007, India}\address{$^{2}$Department of Physics, ARSD College, University of Delhi (South Campus), New Delhi-110021, India}\address{$^{3}$ Max Planck-Institute f\"ur Physik komplexer Systeme, N\"othnitzer Str.38, 01187 Dresden, Germany }

\begin{abstract}
We study the transmission spectra of a Bose Einstein condensate confined in an optical lattice interacting with two modes of a cavity via nonlinear two-photon transition. In particular we show that a nonlinear two-photon interaction between the superfluid (SF) phase and the Mott insulating (MI) phase of a Bose-Einstein condensate (BEC) and the cavity field show qualitatively different transmission spectra compared to the one-photon interaction. We found that when the BEC is in the Mott state, the usual normal mode splitting present in the one-photon transition is missing in the two-photon interaction. When the BEC is in the superfluid state, the transmission spectra shows the usual multiple lorentzian structure. However the separation between the lorentzians for the two-photon case is much larger than that for the one-photon case. This study could form the basis for non-destructive high resolution Rydberg spectroscopy of ultracold atoms or two-photon spectroscopy of a gas of ultracold atomic hydrogen.
\end{abstract}


\maketitle

\section{Introduction}
Cold atoms in optical lattices exhibit phenomena typical of solid state physics like the formation of energy bands, Josephson effects Bloch oscillations and strongly correlated phases. Many of these phenomena have been already the object of experimental investigations. For a recent review see \cite{Morsch06}. Standard methods to observe quantum properties of ultracold atoms are based on destructive matter-wave interference between atoms released from traps \cite{Greiner}. Recently, a new approach was proposed which is based on all optical measurements that conserve the number of atoms. It was shown that atomic quantum statistics can be mapped on transmission spectra of high-Q cavities, where atoms create a quantum refractive index. This was shown to be useful for studying phase transitions between Mott insulator and superfluid states since various phases show qualitatively distinct spectra \cite{Mekhov07}.

Experimental implementation of a combination of cold atoms and cavity QED (quantum electrodynamics) has made significant progress \cite{Nagorny03,Sauer04,Anton05}. Theoretically there have  been some interesting work on the correlated atom-field dynamics in a cavity. It has been shown that the strong coupling of the condensed atoms to the cavity mode changes the resonance frequency of the cavity \cite{Horak00}. Finite cavity response times lead to damping of the coupled atom-field excitations \cite{Horak01}. The driving field in the cavity can significantly enhance the localization and the cooling properties of the system\cite{Griessner04,Maschler04}. It has been shown that in a cavity the atomic back action on the field introduces atom-field entanglement which modifies the associated quantum phase transition \cite{Maschler05,Mekhov07}. The light field and the atoms become strongly entangled if the latter are in a superfluid state, in which case the photon statistics typically exhibits complicated multimodal structures \cite{Chen07}. A coherent control over the superfluid properties of the BEC can also be achieved with the cavity and pump \cite{Bhattacherjee} 

In this work, we show that a nonlinear two-photon interaction between the superfluid (SF) phase and the Mott insulating (MI) phase of a Bose-Einstein condensate (BEC) (confined in an optical lattice) and the cavity field show qualitatively different transmission spectra compared to the one-photon interaction. Two-photon spectroscopy played a very important role in the studies of BEC of atomic hydrogen \cite{fried}. Two-photon excitation of $^{87} Rb$ atoms to a Rydberg state was also achieved recently\cite{low}.

\section{The effective two-photon transition Hamiltonian}

The system we consider here is an ensemble of $N$ two-level atoms with upper and lower states denoted by $|1>$ and $|0>$ respectively in an optical lattice with $M$ sites formed by far off resonance standing wave laser beams inside a cavity. A region of $k\eqslantless M$ sites is coupled to two light modes as shown in Fig.1. Fig.1 shows two cavities containing the two modes $a_{1}$ and $a_{2}$ crossed by a one-dimensional optical lattice confining the BEC. In the two-photon process, an intermediate level $|i>$ is involved, which is assumed to be coupled to $|1>$ and $|0>$ by dipole allowed transitions.
The manybody Hamiltonian in the second quantized form is given by

\begin{figure}[t]
\begin{tabular}{cc}
\includegraphics[scale=0.35]{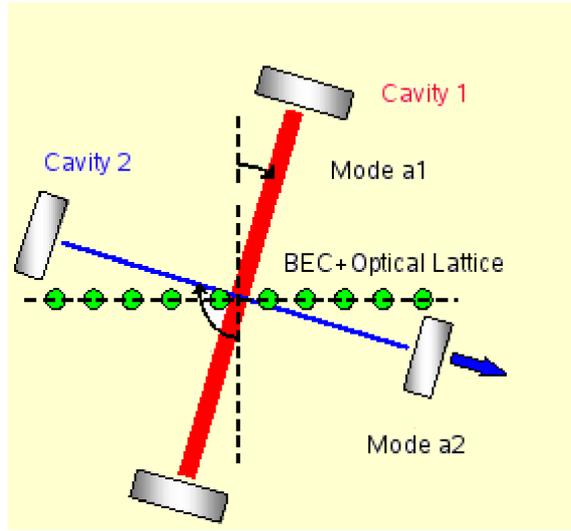}
\end{tabular}

\caption{Schematic diagram of the setup. The BEC atoms are periodically confined in an optical lattice and are made to interact with two laser modes $a_{1}$ and $a_{2}$ which are confined in two intersecting cavities. Mode $a_{2}$ is transmitted and measured by a detector.}
\label{fig.1}
\end{figure}

\begin{subequations}\label{1}
\begin{eqnarray}
H=H_f +H_a, \\
H_f=\sum_{l=1}^{2}{\hbar\omega_l a^\dag_l a_l} -i\hbar\sum_{l=1}^{2}{(\eta^*_l a_l -
\eta_l a^\dag_l)}, \\
H_a=\int{d^3{\bf r}\Psi^\dag({\bf r})H_{a1}\Psi({\bf r})} \nonumber\\
+\frac{2\pi a_s \hbar^2}{m}\int{d^3{\bf r}\Psi^\dag({\bf
r})\Psi^\dag({\bf r})\Psi({\bf r})\Psi({\bf r})}.
\end{eqnarray}
\end{subequations}

In the field part of the Hamiltonian $H_f$, $a_l$ are the annihilation operators of light modes with the frequencies $\omega_l$, wave vectors ${\bf k}_l$, and mode functions $u_l({\bf r})$, which can be pumped by coherent fields with amplitudes $\eta_l$. In the atom part, $H_a$, $\Psi({\bf r})$ is the atomic matter-field operator, $a_s$ is the $s$-wave scattering length characterizing the direct interatomic interaction, and $H_{a1}$ is the atomic part of the single-particle Hamiltonian $H_1$. The detuning between the atomic transition frequency and any one of the two modes is nonzero. Under these circumstances, the intermediate state can be adiabatically eliminated and the effective Hamiltonian of the two-level atom can be written in the rotating-wave and dipole approximation as

\begin{subequations}\label{2}
\begin{eqnarray}
H_1=H_f +H_{a1}, \\
H_{a1}=\frac{{\bf p}^2}{2m_a}+\frac{\hbar\omega_a}{2} \sigma_z -
i\hbar g_0{[\sigma^+ a_1 u_1({\bf r})  a_2  u_2({{\bf r}})-\text{H. c.}}]
\end{eqnarray}
\end{subequations}
Here, ${\bf p}$ and ${\bf r}$ are the momentum and position operators of an atom of mass $m_a$ and resonance frequency $\omega_a$, $\sigma^+$, $\sigma^-$, and $\sigma_z$ are the raising, lowering, and population difference operators, $g_0$ is the atom--light coupling constant assumed to be same for both the modes.

We will consider nonresonant interaction where the light-atom detunings $\Delta = \omega_1+\omega_2 - \omega_a$ are much
larger than the spontaneous emission rate and Rabi frequencies $g_0 a_1 a_2$. Thus, in the Heisenberg equations obtained from the
single-atom Hamiltonian $H_1$ (\ref{2}), $\sigma_z$ can be set to $-1$ (approximation of linear dipoles). Moreover, the polarization
$\sigma^-$ can be adiabatically eliminated and expressed via the fields $a_1$ and $a_{2}$. An effective single-particle Hamiltonian that gives the corresponding Heisenberg equations for $a_1$ and $a_{2}$ can be written as $H_{1\text{eff}}=H_f +H_{a1}$ with 

\begin{equation}\label{3}
H_{a1}=\frac{{\bf p}^2}{2m_a}+V_{\text {cl}}({\bf r})+ \dfrac{2 \hbar g^2_0}{\Delta}a^{\dagger}_{2} a^{\dagger}_{1} a_{1} a_{2} |u_{1}({\bf r})|^{2} |u_{2}({\bf r})|^{2}.  
\end{equation}

Here, we have also added a classical trapping potential of the lattice, $V_{\text {cl}}({\bf r})$, corresponds to a strong classical standing wave. Interestingly, we find that unlike in \cite{Mekhov07} we find that the Hamiltonian $H_{a1}$ does not contain terms like $u^{*}_{1}({\bf r})u_{2}({\bf r})$ or $u^{*}_{2}({\bf r})u_{1}({\bf r})$ which gives rise to an optical grating.

To derive the generalized Bose--Hubbard Hamiltonian we expand the field operator $\Psi({\bf r})$ in Eq.~(\ref{1}), using localized Wannier functions corresponding to $V_{\text {cl}}({\bf r})$ and keeping only the lowest vibrational state at each site: $\Psi({\bf
r})=\sum_{i=1}^{M}{b_i w({\bf r}-{\bf r}_i)}$, where $b_i$ is the annihilation operator of an atom at the site $i$ with the coordinate
${\bf r}_i$. Substituting this expansion in Eq.~(\ref{1}) with $H_{a1}$ (\ref{3}), we get 

\begin{eqnarray}\label{4}
H=H_f+\sum_{i,j=1}^M{J_{i,j}^{\text {cl}}b_i^\dag b_j} + \dfrac{2 \hbar g^2_0}{\Delta} a^{\dagger}_{2}a^{\dagger}_{1}a_{1}a_{2} \sum_{i,j=1}^{K} J_{i,j}b^{\dagger}_{i} b_{j}   \nonumber \\
+\frac{U}{2}\sum_{i=1}^M{b_i^\dag b_i(b_i^\dag b_i-1)},
\end{eqnarray}

where 

\begin{equation}\label{5}
J_{i,j}^{\text {cl}}=\int{d{\bf r}}w({\bf r}-{\bf
r}_i)\left(-\frac{\hbar^2\nabla^2}{2m}+V_{\text {cl}}({\bf
r})\right)w({\bf r}-{\bf r}_j),
\end{equation}

\begin{equation}\label{6}
J_{i,j}=\int{d{\bf r}}w({\bf r}-{\bf r}_i) |u_{1}({\bf r})|^{2} |u_{2}({\bf r})|^{2}  w({\bf r}-{\bf r}_j),
\end{equation}

\begin{equation}
U=4\pi a_s\hbar^2/m_a \int{d{\bf r}|w({\bf
r})|^4}.
\end{equation}

The BH Hamiltonian derived above is valid only for weak atom-field nonlinearity \cite{larson}. We assume that atomic tunneling is possible only to the nearest neighbor sites. Thus, coefficients (\ref{5}) do
not depend on the site indices ($J_{i,i}^{\text {cl}}=J_0^{\text {cl}}$ and $J_{i,i\pm 1}^{\text {cl}}=J^{\text {cl}}$), while
coefficients (\ref{6}) are still index-dependent. The Hamiltonian (\ref{4}) then reads 

\begin{eqnarray}\label{7} 
H=H_f+J_0^{\text {cl}}\hat{N}+J^{\text {cl}}\hat{B}+ \dfrac{2 \hbar g^2_0}{\Delta}
 a^{\dagger}_{2} a^{\dagger}_{1} a_{1} a_{2} \left(\sum_{i=1}^K{J_{i,i} \hat{n}_i}\right)  \nonumber \\
+ \dfrac{2 \hbar g^2_0}{\Delta} a^{\dagger}_{2} a^{\dagger}_{1} a_{1} a_{2} \left(\sum_{<i,j>}^K{J_{i,j} b_i^\dag
b_j}\right) +\frac{U}{2}\sum_{i=1}^M{\hat{n}_i(\hat{n}_i-1)},
\end{eqnarray}

where $<i,j>$ denotes the sum over neighboring pairs, $\hat{n}_i=b_i^\dag b_i$ is the atom number operator at the $i$-th site, and $\hat{B}=\sum_{i=1}^M{b^\dag_i b_{i+1}}+{\text {H.c.}}$ While the total atom number determined by $\hat{N}=\sum_{i=1}^M{\hat{n}_i}$ is conserved, the atom number at the illuminated sites, determined by $\hat{N}_K=\sum_{i=1}^K{\hat{n}_i}$, is not necessarily a conserved
quantity.

\section{The Transmission Spectra}

The Heisenberg equations for $a_1$, $a_2$ and $b_i$ can be obtained from
the Hamiltonian (\ref{7}) as

\begin{equation} \label{8}
\dot a_{1}=-i\left\lbrace \omega_{1} +2 \dfrac{g_{0}^{2}}{\Delta}(\sum_{i=1}^{K}J_{i,i}\hat{n}_{i}+\sum_{<i,j>}^{K} J_{i,j}b^{\dagger}_{i} b_{j}) a^{\dagger}_{2} a_{2}\right\rbrace a_{1}+\eta_{1} 
\end{equation}

\begin{equation} \label{9}
\dot a_{2}=-i\left\lbrace \omega_{2} +2 \dfrac{g_{0}^{2}}{\Delta}(\sum_{i=1}^{K}J_{i,i}\hat{n}_{i}+\sum_{<i,j>}^{K} J_{i,j}b^{\dagger}_{i} b_{j}) a^{\dagger}_{1} a_{1}\right\rbrace a_{2}+\eta_{2} 
\end{equation}

\begin{eqnarray}\label{10}
\dot{b}_i=-\frac{i}{\hbar}\left( J_0^{\text {cl}}+ \dfrac{2 \hbar g^2_0}{\Delta} a^{\dagger}_{2} a^{\dagger}_{1} a_{1} a_{2} J_{i,i}+U\hat{n}_i\right) b_i  \nonumber \\
-\frac{i}{\hbar}\left( J^{\text {cl}}+\dfrac{2 \hbar g^2_0}{\Delta} a^{\dagger}_{2} a^{\dagger}_{1} a_{1} a_{2} J_{i,i+1}\right)b_{i+1} \nonumber \\
-\frac{i}{\hbar}\left( J^{\text {cl}}+ \dfrac{2 \hbar g^2_0}{\Delta} a^{\dagger}_{2} a^{\dagger}_{1} a_{1} a_{2}  J_{i,i-1}\right)b_{i-1}.
\end{eqnarray}

Equations \ref{8}-\ref{10} represent the coupled light-matter wave equations which determine completely the dynamics of the present system. In Eqns. \ref{9} and \ref{10}, the second and third terms in the parentheses correspond to the phase shift of the light modes $a_{1}$ and $a_{2}$ due to two-photon coherence.  Note that the term that describes scattering of one mode into the other is absent. We consider a deep lattice formed by a strong classical potential $V_{\text {cl}}({\bf r})$, so that the overlap between Wannier functions in Eqs.~(\ref{5}) and (\ref{6}) is small. Thus, we can neglect the contribution of tunneling by putting $J^{\text{cl}}=0$ and $J_{i,j}=0$ for $i \ne j$. Under this approximation, the matter-wave dynamics is not essential for light scattering. In experiments, such situation can be realized because the time scale of light measurements can be much faster than the time scale of atomic tunneling. One of the well-known advantages of the optical lattices is their extremely high tunability. Thus, tuning the lattice potential, tunneling can be made very slow~\cite{jaksch}.
In a deep lattice, the on-site coefficients $J_{i,i}$ (\ref{6}) can be approximated as $J_{i,i}=|u_1({\bf r}_i)|^{2} |u_2({\bf r}_i)|^{2}$ neglecting details of the atomic localization. Using $a_{i}(t)=a_{i} exp(-i\omega_{1}t)$, we obtain the stationary solutions of Equations \ref{8} and \ref{9} as

\begin{equation}\label{11}
a^{\dagger}_{1} a_{1}=\dfrac{\eta_{1}^{2}}{\left\lbrace \dfrac {2 g_{0}^{2}}{\Delta} \sum_{i=1}^{K} |u_{1}({\bf r})|^{2} |u_{2}({\bf r})|^{2} \hat {n}_{i} a^{\dagger}_{2} a_{2}\right\rbrace^{2}+\kappa^{2}} 
\end{equation}

\begin{equation}\label{12}
a^{\dagger}_{2} a_{2}=\dfrac{\eta_{2}^{2}}{\left\lbrace \Delta_{p}-\dfrac{ 2 g_{0}^{2}}{\Delta} \sum_{i=1}^{K} |u_{1}({\bf r})|^{2} |u_{2}({\bf r})|^{2} \hat {n}_{i} a^{\dagger}_{1} a_{1}\right\rbrace^{2}+\kappa^{2}} 
\end{equation}

Where $\Delta_{p}=\omega_{1}-\omega_{2}$. We now assume the mode $a_{1}$ to be in a the coherent state, which enables us to consider the quantity $a_{1}$ as a $c$ number. The intensity of the mode $a_{1}$ is then large and undepleted as compared to the weak mode ($a_{2}$). As a result in Eqn. \ref{11}, one could neglect the influence of the mode $a_{2}$ on $a_{1}$ by ignorning the quantity $\dfrac{g_{0}^{2}}{\Delta} \sum_{i=1}^{K} |u_{1}({\bf r})|^{2} |u_{2}({\bf r})|^{2} \hat {n}_{i} a^{\dagger}_{2} a_{2}$ with respect to $\kappa$. This basically means that we are ignorning the phase shift of the mode $a_{1}$. This can be achieved by keeping the detuning $\Delta$ large and the probe amplitude $\eta_{2}$ small. This yields

\begin{equation}\label{13}
a^{\dagger}_{2} a_{2}=\dfrac{\eta_{2}^{2}}{\left\lbrace \Delta_{p}-\dfrac{2 g_{0}^{2} \eta_{1}^{2}}{\kappa^{2} \Delta } \sum_{i=1}^{K} |u_{1}({\bf r})|^{2} |u_{2}({\bf r})|^{2} \hat {n}_{i} \right\rbrace^{2}+\kappa^{2}} 
\end{equation}

Following \cite{Mekhov07} Eq.~(\ref{13}) allows to express  $a^{\dagger}_{2} a_{2}$ as a function $f(\hat{n}_1,...,\hat{n}_M)$ of atomic occupation number operators and calculate their expectation values for prescribed atomic states $|\Psi\rangle$. 

For the Mott state  $\langle\hat{n}_i\rangle_\text{MI}=q_i$ atoms are well localized at the $i$th site with no number fluctuations. It is represented by a product of Fock states, i.e. $|\Psi\rangle_\text{MI}=\prod_{i=1}^M |q_i\rangle_i\equiv |q_1,...,q_M\rangle$, with expectation values

\begin{eqnarray}\label{14}
\langle f(\hat{n}_1,...,\hat{n}_M)\rangle_\text{MI}=f(q_1,...,q_M),
\end{eqnarray}

For simplicity we consider equal average densities $\langle\hat{n}_i\rangle_\text{MI}=N/M\equiv n$ ($\langle\hat{N}_K\rangle_\text{MI}=nK\equiv N_K$).

In SF state, each atom is delocalized over all sites leading to local number fluctuations. It is represented by superposition of Fock states corresponding to all possible distributions of $N$ atoms at $M$ sites: $|\Psi\rangle_\text{SF} =\sum_{q_1,...,q_M}\sqrt{N!/M^N}/\sqrt{q_1!...q_M!} |q_1,...,q_M\rangle$.  Expectation values of light operators can be calculated from 

\begin{eqnarray}\label{15}
\langle f(\hat{n}_1,...,\hat{n}_M)\rangle_\text{SF}=\frac{1}{M^N}
\sum_{q_1,...,q_M}\frac{N!} {q_1!...q_M!}f(q_1,...,q_M),
\end{eqnarray}

representing a sum of all possible ``classical'' terms. Thus, all these distributions contribute to scattering from a SF, which is
obviously different from $\langle f(\hat{n}_1,...,\hat{n}_M)\rangle_\text{MI}$ with only a single contributing term.
We now apply this formalism to compare the two-photon spectra with the one photon spectra calculated in \cite{Mekhov07} Eqn. (5). For travelling waves, we take $|u_{1}({\bf r_{i}})|^{2}=|u_{2}({\bf r_{i}})|^{2}=1$ and $\sum_{i=1}^{K}|u_{1}({\bf r_{i}})|^{2} |u_{2}({\bf r_{i}})|^{2}\hat{n}_{i}=\sum_{i=1}^{K} \hat{n}_{i}$.

\begin{figure}[t]
\begin{tabular}{cc}
\includegraphics[scale=0.8]{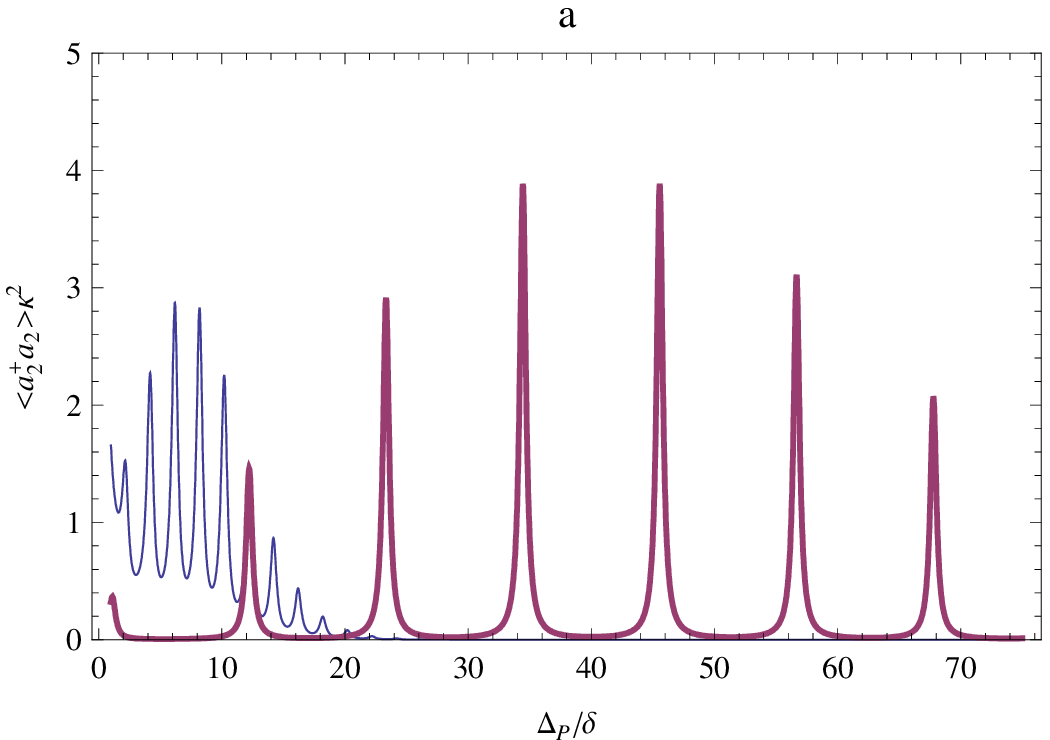}& \includegraphics[scale=0.8]{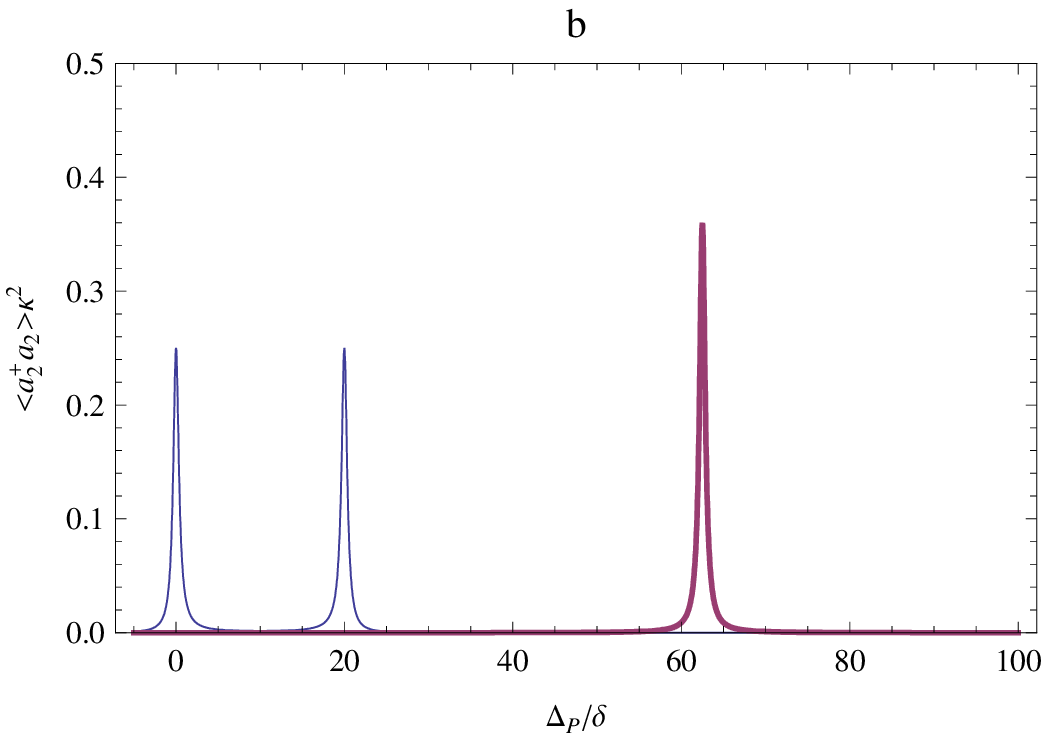}
\end{tabular}

\caption{Photon number in the mode $a_{2}$ for SF state (a) and MI state (b) due to one-photon transition (thin line) and two-photon transition (thick line). The separation between the lorentzians in the SF state due to two-photon case is $11 \delta$ and that due to one-photon case is $2 \delta$. In the Mott state for the one-photon case, two-satellite contour is observed reflecting normal mode splitting of the two oscillators $<a_{0,1}>$ coupled through atoms. This splitting is absent in the two-photon case. All the frequencies are scaled with respect to $\delta$. Here $\delta=g_{0}^{2}/\Delta$. Also $N=M=K=15$ and $\kappa=0.3 \delta$. The mode amplitudes are in scaled units $\eta_{1}=1$ and $\eta_{2}=0.6$. }
\label{fig.2}
\end{figure}

\begin{figure}[t]
\begin{tabular}{cc}
\includegraphics[scale=0.8]{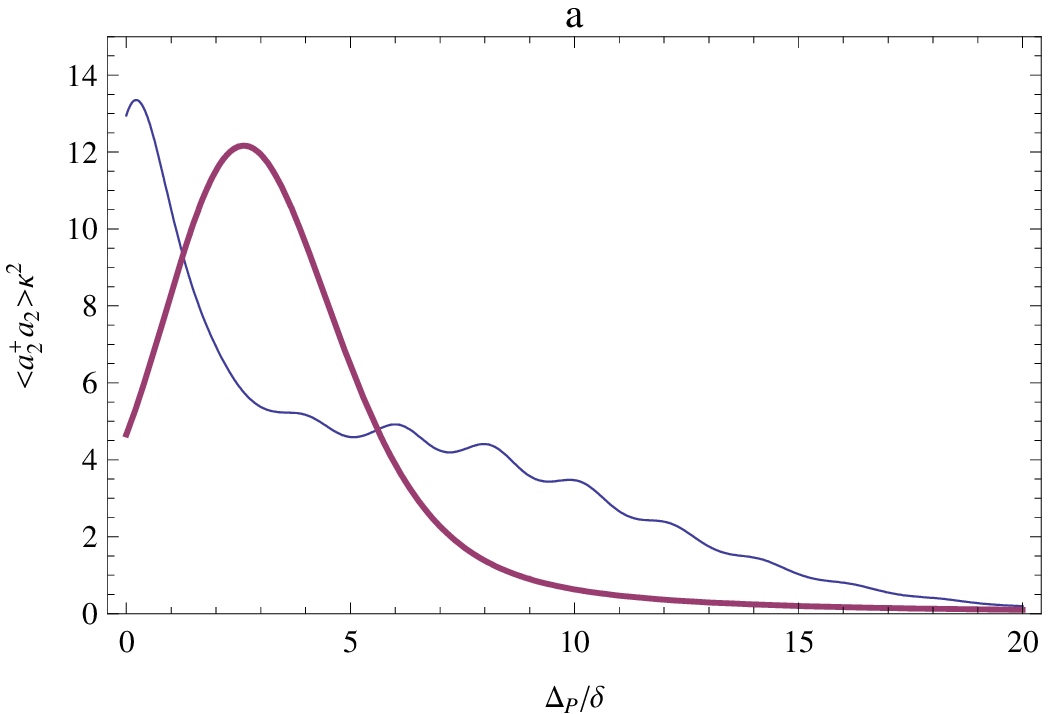}& \includegraphics[scale=0.8]{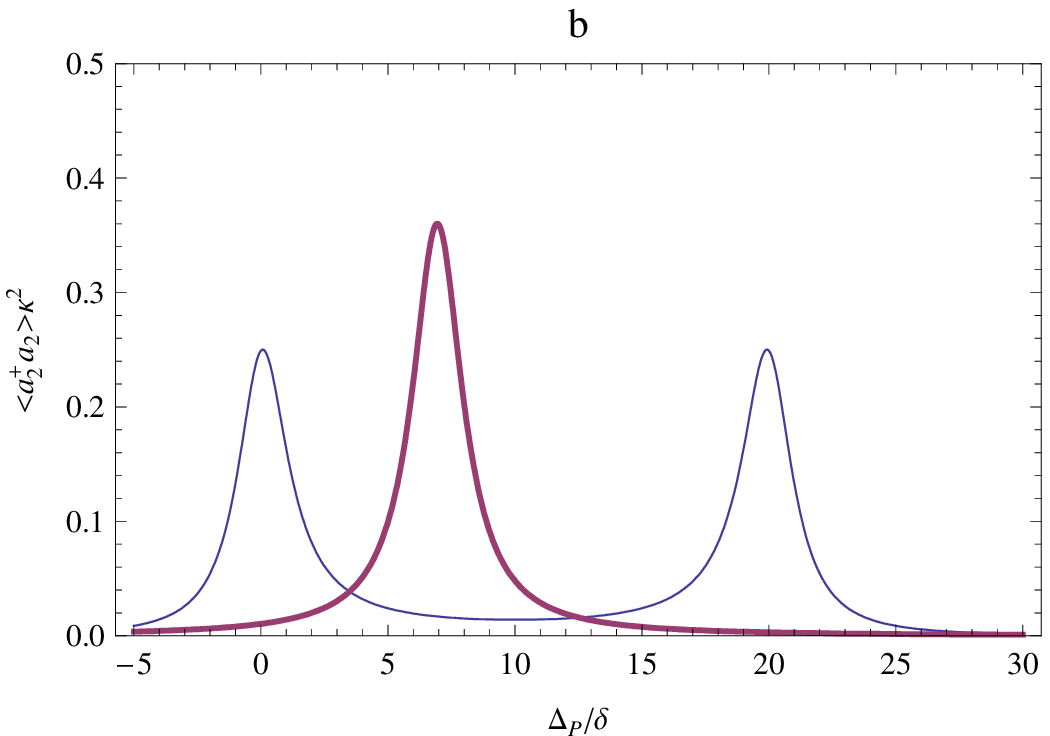}
\end{tabular}

\caption{ Photon number in the mode $a_{2}$ for SF state (a) and MI state (b) due to one-photon transition (thin line) and two-photon transition (thick line). The parameters are same as in Fig. 2 except $\kappa=1.2 \delta$.  The comb like structure for the SF case (a) is replaced by a broad lorentzian for both the single-photon and two-photon case.  For the Mott case (b), however, the peaks are broadened and the two-photon single peak shifts i.e, the dispersion is reduced. }
\label{fig.3}
\end{figure}

We present the transmission spectra for the one-photon case(thin line) and two photon (thick line) case in the SF state in Fig.1(a) and that in the Mott state in Fig.1(b) for the case where $\kappa=0.3 g^{2}_{0}/\Delta$. Clearly for both the cases, the transmission spectra is a sum of lorentzians with different dispersion shifts. A comb like structure is seen if each lorentzian is resolved. However the separation between the lorentzians for the two-photon case ($11 \delta$) is much larger than that for the one-photon case($2 \delta$).Such a spectra can be reproduced experimentally by repeated measurements over a long time scale so that a superposition of different atomic distributions(obtained by tunneling) contibutes to the spectra. In the Mott state, the difference between the one-photon interaction(thin line) and the two-photon interaction (thick line) is striking. In the one-photon case, two-satellite contour is observed reflecting normal mode splitting of the two oscillators $<a_{0,1}>$ coupled through atoms. This splitting is absent in the two-photon case. A probable explanation for this observation can be traced back to the Hamiltonian $H_{a1}$ of Eqn. \ref{2}. In the absence of any optical grating the atomic grating is not formed and as a result diffraction of one mode into the other is absent (absence of Bragg scattering). In the one photon case, the two modes are bahaving like two independent oscillators exchanging energy via the atoms. In the two-photon transition, there is no energy exchange between the two modes but energy exchange occurs between the atoms and the two modes (taken together). The system then behaves effectively as a two-level atom interacting with a single mode via one-photon transition.
The results for the case $\kappa > g^{2}_{0}/\Delta$ is shown in Fig.2. As found in \cite{Mekhov07} , the comb like structure for the SF case is replaced by a broad lorentzian for both the single-photon and two-photon case (Fig2a).  For the Mott case, however, the peaks are broadened and the two-photon single peak shifts i.e, the dispersion is reduced.

From the experimental point of view, such a two photon excitation has already been achieved in $^{87}Rb$ atoms \cite{low} and atomic hydrogen BEC \cite{fried}. Using the setup described in \cite{low}, $^{87}Rb$-atoms are magnetcally trapped in the $5S_{1/2}, F=2, m_{F}=2$ state and produce samples from thermal clouds to BECs by means of forced rf evaporation. After this preparation, the atoms are subject to a two-photon Rydberg excitation (using $780.246 nm$ and $480.6nm$ lasers) via the $5P_{3/2}$ state to the $43S_{1/2}$ state with a length of the square light pulses between $170 ns$ and $2\mu s$. To reduce spontaneous photon scattering, the light is blue detuned by $2 \pi\times 483 MHz$ from the $5P_{3/2}, F=3$ level. Thus only one photon per 100 atoms is scattered for the longest excitation time. However in such experiments the excitation is locally blocked by the van der Walls interaction between Rydberg atoms.

\section{Conclusion}

We have investigated off-resonant collective light scattering from a Bose Einstein condensate trapped in an optical lattice interacting with two cavity modes via nonlinear two-photon transition. Measuring the transmission spectra allows one to distinguish between a nonlinear two-photon interaction and a one-photon interaction. Depending on the state of the BEC, we found that the two-photon spectra is different from one-photon spectra. We found that when the BEC is in the Mott state, the usual normal mode splitting present in the one-photon transition is missing in the two-photon interaction. When the BEC is in the superfluid state, the transmission spectra shows the usual multiple lorentzian structure. However the separation between the lorentzians for the two-photon case is much larger than that for the one-photon case. This method could be useful for non-destructive high resolution Rydberg spectroscopy of ultracold atoms or two-photon spectroscopy of hydrogen BEC. 

\section{Acknowledgements}

One of the authors Tarun Kumar acknowledges the Council of University Grants Commission, New Delhi for the financial support under the Junior Research Fellowship scheme Sch/JRF/AA/30/2008-2009.

\end{document}